\documentclass[useAMS,usenatbib]{mn2e} 
\usepackage{epsfig}
\usepackage{graphicx,amsmath,multicol}
\usepackage{psfrag}
\usepackage[dvipsnames]{xcolor}
\definecolor{webgreen}{rgb}{0,.5,0}
\definecolor{webbrown}{rgb}{.6,0,0}
\usepackage[pdfpagelabels]{hyperref}
\usepackage[normalem]{ulem}
\newcommand {\tr} {t_{\rm {cool}} / t_{\rm {ff}}}
\defcitealias{ps05}{PS05}
\hypersetup{%
   colorlinks=true,hyperfootnotes=false,%
   breaklinks=true,%
   plainpages=false, bookmarksnumbered, bookmarksopen=true,
   bookmarksopenlevel=1,%
   urlcolor=webbrown, linkcolor=RoyalBlue, citecolor=webgreen,%
}

\newcommand{\comments}[1]{}
\newcommand       \be           {\begin{equation}}
\newcommand       \ee           {\end{equation}}
\newcommand       \ba           {\begin{eqnarray}}
\newcommand       \ea           {\end{eqnarray}}

\def\msun{\rm \ M_\odot}

\def\lesssim{\mathrel{\hbox{\rlap{\hbox{\lower4pt\hbox{$\sim$}}}\hbox{$<$}}}}
\def\gtrsim{\mathrel{\hbox{\rlap{\hbox{\lower4pt\hbox{$\sim$}}}\hbox{$>$}}}}

\title[Evolution of density perturbations in the ICM]
{The cold mode: A phenomenological model for the evolution of density perturbations in the intracluster medium}  

\author[A. Singh, P.\ Sharma]
{Ashmeet Singh $^\dag$, Prateek Sharma$^\ddag$\\
$^\dag$ Department of Physics, Indian Institute of Technology Roorkee, Roorkee 247667, Uttarakhand, India (ashmtuph@iitr.ac.in)\\
$^\ddag$Department of Physics and Joint Astronomy Program, Indian Institute of Science, Bangalore, India 560012 (prateek@physics.iisc.ernet.in)}

\begin{document}

\pagerange{\pageref{firstpage}--\pageref{lastpage}} \pubyear{2012}
\maketitle

\label{firstpage}

\begin{abstract}
Cool cluster cores are in global thermal equilibrium but are locally thermally unstable.
We study a nonlinear phenomenological model for the evolution of density perturbations in the ICM due to local thermal instability and gravity.
We have analyzed and extended a model for the evolution of an over-dense blob in the ICM. 
We find two regimes in which the over-dense blobs can cool to thermally stable low temperatures. 
One for large $t_{{\rm {cool}}} / t_{\rm {ff}}$ ($t_{{\rm {cool}}}$ is the cooling time and $t_{{\rm {ff}}}$ is the free fall time), where a large initial 
over-density is required for thermal runaway to occur; this is the regime which 
was previously analyzed in detail. We discover a second regime for 
$t_{\rm {cool}} / t_{\rm {ff}} \lesssim 1$ (in agreement with Cartesian simulations of local thermal instability in an external gravitational field), where 
runaway cooling happens for arbitrarily small amplitudes. Numerical simulations have shown that cold gas condenses out more easily in a spherical geometry. 
We extend 
the analysis to include geometrical compression in weakly stratified atmospheres such as the ICM. 
With a single parameter, analogous to the mixing length, we are able to reproduce the 
results from numerical simulations; namely, small density perturbations lead to the condensation of extended cold filaments only if $\tr \lesssim 10$. 
\end{abstract}


\section{Introduction}           
\label{sect:intro}
The intracluster medium (ICM) of galaxy clusters is composed of ionized plasma at the virial temperature ($\sim 10^7-10^8$ K). The ICM is 
heated due to the release of gravitational energy which is converted into thermal energy at the accretion shock. Given the ICM temperature 
(\citealt{mo99}), the lighter elements like H and He are fully ionized. The typical density in cluster cores lines in the range of $0.001-0.1$ cm$^{-3}$ 
(e.g., \citealt{accept09}). The hot ionized ICM emits X-Rays due to free-free and bound-free/bound-bound emission, which is enhanced due to the 
high metallicity ($\sim 0.3$ solar) of the ICM  (\citealt{ryb+light}). The cooling loss rate per unit volume is proportional to the square of the particle 
number density. Since the inner ICM is denser than the outer regions, the gas near the center is expected to cool faster than the gas in the 
outskirts. This relatively fast cooling of the gas near the center of the cluster should reduce the thermal support it provides to the overlying 
layers. Hence, the dilute outer layers are expected to flow into the inner cooling region, resulting in a subsonic flow of the ICM plasma called 
the cooling flow. The cooling flow, being subsonic, maintains quasi-hydrostatic equilibrium in the ICM (\citealt{fabian94}). 

Observations of galaxy clusters by X-Ray satellites such as  {\it Chandra} and {\it XMM-Newton} (e.g., \citealt{peterson2003}) show a dramatic 
lack of cooling. Similarly, signs of cold gas and star-formation are missing (e.g., \citealt{edge2001,odea2008}), contrary to the predictions of the 
cooling flow model. It is therefore indicative, that some form of heating in the ICM balances cooling and explains the lack of a cooling flow. The 
ICM, thus, stays in a state of global thermal equilibrium where heating, on average, balances radiative cooling. Of the various possibilities discussed, 
heating due to mechanical energy injection by jets and bubbles driven by the central Active Galactic Nuclei (AGN) appears to be the most promising 
(see \citealt{mcnamara2007} for a review).
  
\citet{ps05}, and later \citet{ps10}, have argued that over-dense blobs of gas cool faster than the rest of the ICM. These fast-cooling blobs become 
heavier than the background ICM, sink and feed the central Active Galactic Nucleus (AGN). This cold feedback from non-linear perturbations 
triggers winds and/or jets from the AGN which heat the ICM and maintain the global thermal equilibrium, as mentioned earlier. The mechanical 
energy input creates density perturbations in the ICM, which if the ICM is sufficiently dense, can lead to the next episode of multiphase cooling 
and feedback heating. 

While \citealt{ps05} focus on the evolution of density perturbations, the aim of \citealt{ps10} is to show that the cold blobs can lose angular 
momentum sufficiently fast due to drag and cloud-cloud collisions. In the current paper we are not concerned about the angular momentum 
problem but about the condensation of cold gas from almost spherical, non-rotating ICM. Later works, with various degrees of realism (e.g., \citealt{sharma12,gaspari12,li14}), have vindicated this basic picture in which condensation and accretion of cold gas plays a key role in 
closing the feedback loop.
 
 \citealt{ps05} (hereafter PS05) consider a Cartesian setup, not accounting for geometric compression due to radial gravity. We expect geometrical 
 compression to be important if over-dense blobs travel a distance comparable to the location of their birth. The differences between plane-parallel 
 and spherical geometries have been highlighted by \citealt{mccourt12} and \citealt{sharma12}.\footnote{The plane-parallel and spherical atmospheres 
 are physically distinct; they are not the same initial conditions solved using different coordinate systems. In a plane-parallel atmosphere gravitational 
 field vectors are everywhere parallel but in a spherical atmosphere the gravitational field points toward a single point.} 
 They show that the cold blobs seeded by small 
 perturbations saturate at large densities (and thermally stable temperatures) if $t_{\rm cool}/t_{\rm ff} \lesssim 1$ in a plane-parallel atmosphere 
 and if $t_{\rm cool}/t_{\rm ff} \lesssim 10$ in a spherical atmosphere; i.e., cold gas condenses more easily in a spherical atmosphere. The ratio 
 $\tr$ of the background ICM (more precisely, $t_{\rm TI}/t_{\rm ff}$, the ratio of thermal instability timescale and the free-fall time) is an important 
 parameter governing the evolution of small density perturbations. For small $\tr$ the cooling time is short and a slightly over-dense blob cools
 to the stable temperature and falls in ballistically. On the other hand, for large $\tr$, the over-dense blob responds to gravity as it is cooling; 
 the shear generated due to the motion of the over-dense blob relative to the ICM leads to mixing of the cooling blob and it cannot cool to the 
 stable temperature.
 
 In this paper we extend PS05's analysis to account for spherical compression as the blob travels large distances. In order that spherical 
 compression does not over-compress blobs, we also include a model to account for the mixing of blobs. With a single adjustable 
 parameter, we are able to reproduce the results of \citet{sharma12} -- that cold gas condenses with tiny perturbations if 
 $t_{\rm cool}/t_{\rm ff} \lesssim 10$ -- for a wide range of cluster parameters. In addition to this, we also highlight that there are two 
 regimes for the evolution of over-dense blobs, in both Cartesian and spherical geometries: for  $t_{\rm cool}/t_{\rm ff}$ smaller than a 
 critical value ($\approx 1$ in a plane-parallel and $\approx 10$ in a spherical atmosphere) the over-dense blobs cool in a runaway 
 fashion; for $t_{\rm cool}/t_{\rm ff}$ greater than the critical value a finite amplitude is needed (a larger amplitude is required for a 
 higher $t_{\rm cool}/t_{\rm ff}$) for runaway cooling. PS05 have highlighted the second point and have argued that nonlinear amplitudes 
 are required for producing cold gas in cluster cores. However, as \citet{mccourt12, sharma12}  show, with spherical compression even 
 small amplitude perturbations should give cold gas in cool-core clusters.
 
We emphasize that our paper presents a simple analytic model for the evolution of density perturbations in the ICM. We are interested in 
the qualitative physics of the evolution of over-dense blobs in cluster cores and in comparing analytic models with numerical simulations. 
Quantitative results with accurate modeling of nonlinear processes such as mixing can only be obtained via numerical simulations. 
Moreover, our analytic models neglect effects such as thermal conduction, cosmic rays, magnetic fields, turbulence. Examples of recent 
simulations of the interplay of cooling and AGN feedback, including some of these effects, are 
\citet{dubois11,sharma12,gaspari12,li14,wagh14,banerjee14}. 
 
The paper is organized as follows. In section 2 we present the 1-D profiles of the ICM (density, temperature, etc.) in hydrostatic 
equilibrium with a fixed dark matter halo. In section 3 we describe the equations governing the evolution of an over-dense blob in 
the ICM. In section 4 we present our results, starting with our phenomenological model to account for geometrical compression. 
We also discuss blob evolution for different cluster and blob parameters, contrasting the evolution in Cartesian and spherical 
atmospheres. In section 5 we conclude with astrophysical implications.

\section{1D model for the intracluster medium}

Hydrostatic equilibrium (HSE) in the ICM is a good approximation for the relaxed, cool-core clusters (for observational constraints 
on turbulent pressure support, see \citealt{churazov08,werner09}). Thus,
\begin{equation}
\label{eq:hse}
\frac{dP}{dr} = -\rho(r) g(r)	\: ,
\end{equation}
where {\it P} is the gas pressure, $\rho$ is the gas mass density and {\it g} is the acceleration due to gravity in the cluster at a 
distance {\it r} from the center.

The Navarro-Frenk-White (\citealt{nfw96}) (NFW) profile provides a good  description of the dark matter (DM) distribution in relaxed 
halos such as galaxy clusters. We have used a spherically symmetric NFW profile for the dark matter as:
\begin{equation}
\label{eq:NFW}
\frac{\rho_{\rm DM} (r)}{\rho_{\rm crit}} = \frac{\Delta_{c}}{{r/r_s} (1 + [r/r_s]^{2})}	,
\end{equation}
where $\Delta_{c}$ and $r_{s}$ are the characteristic density parameter and the scale radius, and $\rho_{\rm crit}$ is the critical 
density of the universe. Generally the size of the dark matter halo is taken to be $r_{200}$, the virial radius around the cluster center 
within which the average dark matter density is 200 times the critical density of the universe. The NFW parameters can be recast 
in terms of the concentration parameter $c = r_{200}/r_{s}$, where
\begin{equation}
\label{DeltacNFW}
\Delta_{c} = \frac{200}{3} \frac{c^{3}}{\ln (1+c) - c/(1+c)}	\:	.
\end{equation}
 The corresponding acceleration due to gravity in the ICM is
\begin{equation}
\label{eq:gNFW}
g_{\rm NFW}(r) = \frac{G M_{\rm DM,encl}(r)}{r^{2}}	\: ,
\end{equation}
where we ignore the gravity due to baryons as their mass fraction is small and they are more diffuse compared to dark matter. The 
cluster mass is given by $M_{\rm DM,200}$, the dark matter mass enclosed within the sphere of radius $r_{200}$.

The ionized plasma (most of the baryonic matter) in clusters is confined by the gravitational potential of the dark matter. To study the evolution 
of over-dense blobs in a generic ICM, we numerically compute the ICM profile by solving hydrostatic equilibrium in the NFW 
gravitational potential (Eq. \ref{eq:gNFW}) and by specifying the entropy profile of the ICM. The X-ray entropy profile for the 
ICM, $K$ (in keV cm$^{2}$) is defined to be the adiabatic invariant (for temperature in keV $T_{\rm keV}$, electron number 
density $n_{e}$, adiabatic index $\gamma = \frac{5}{3}$)
\begin{equation}
\label{eq:entropyIndex}
K = \frac{T_{\rm keV}}{n_{e}^{\gamma - 1}}  \: .
\end{equation}
The entropy profile of most galaxy clusters is reasonably well fit by a model given by,
\begin{equation}
\label{eq:entropyChandra}
K = K_{0} \: + \: K_{100} \: \left( \frac{r}{100 \rm kpc} \right) ^{\alpha} \:,
\end{equation}
where the parameters $K_{0}, K_{100}$ and $\alpha$ are the parameters introduced by \citet{accept09}. Combining this with the 
hydrostatic equilibrium equation yields for the total particle number density {\it n},
\begin{equation}
\label{eq:dndr}
\frac{dn}{dr} = -\frac{n}{\gamma \: K} \left[ \frac{{\mu_{e}}^{\gamma - 1} m_{p} \: g_{\rm NFW} \: n^{1 - \gamma}}{{\rm keV} k_{B} \mu^{\gamma - 2}} + \frac{dK}{dr} \right] \:,
\end{equation}
where keV=$1.16\times 10^7$ (the conversion of keV to Kelvin), $m_{p}$ is the proton mass, and $\mu$/$\mu_{e}$/$\mu_{i}$ are 
the mean mass per particle/electron/ion, governed by $\rho = n \mu m_{p} = n_{e} \mu_{e} m_{p} = n_{i} \mu_{i} m_{p}$. The total 
number density is the sum of electron and ion number densities $n = n_{e} + n_{i}$. We choose a constant ICM metallicity 
corresponding to the standard value of a third of the solar value (\citealt{leccardi08}); thus, $\mu = 0.62$ and $\mu_{e} = 1.17$. 
From Eq.~\ref{eq:entropyChandra}, we can get $dK/dr$ and by specifying the number density at a point in the ICM as a boundary 
condition, ICM profiles for pressure, density, etc. can be solved for.
The entropy profile of the ICM from Eq.~\ref{eq:entropyChandra} is a monotonically increasing function of {\it r} and hence the 
ICM is convectively stable according to the Schwarzschild criterion for adiabatic atmospheres.\footnote{The Schwarzschild 
criterion is not applicable for the magnetized ICM plasma with anisotropic conduction, and the ICM core is unstable to the 
heat-flux driven buoyancy instability (\citealt{quataert08}). However, this instability saturates by reorienting field lines 
perpendicular to the radial direction and in the saturated state the response of a perturbed blob is stable, qualitatively similar 
to a convectively stable adiabatic atmosphere (\citealt{sharma09}). Therefore, we ignore the effects of thermal conduction 
in this paper.} An adiabatic blob will oscillate in the atmosphere with the local Brunt-$\rm V\ddot{a}is\ddot{a}l\ddot{a}$ oscillation 
frequency (c.f. Eq. \ref{eq:BVOscillation}). However, as we show later, spherical compression is important for weakly stratified
medium such as the ICM.

\begin{figure}
  \includegraphics[scale=0.35]{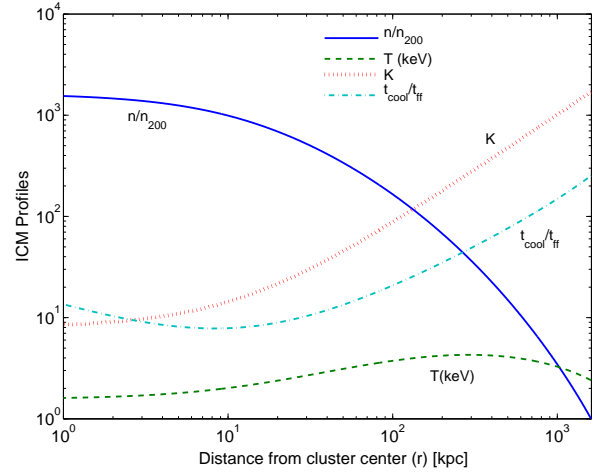}
  \centering
  \caption{Various profiles  for the fiducial cluster (see Table \ref{profile_table}): normalized gas number density $(n/n_{200})$, 
  temperature $T$ in keV, entropy $K~({\rm keV cm}^2)$ and the ratio of the cooling time and the free-fall time ($t_{\rm{cool}}/t_{\rm{ff}}$).}\label{fig:fiducial}
\end{figure}

In Figure~\ref{fig:fiducial}, we plot the profiles of different variables with NFW parameters $c = 3.3$ and $M_{200} = 5.24 \times 10^{14} M_{\odot}$, 
and with entropy parameters $K_{0} = 8~{\rm keV cm^{2}}, K_{100} = 80~\rm{keV cm^{2}}, \alpha = 1.1$ and $n(r_{200}) \: = 10^{-4} \: {\rm cm^{-3}}$.  
These parameters correspond to the fiducial run listed in Table \ref{profile_table}. Our fiducial cluster has a virial radius $r_{200} \approx 1603 \: \rm{kpc}$ 
and $M_{\rm gas}/M_{200}$ (hot gas to DM mass ratio) = 0.18. While we use the same NFW halo parameters throughout the paper (except in Fig. 
\ref{fig:lower_mass}, where we investigate the blob evolution in a $10^{12} \msun$ halo), we use different 
entropy parameters for investigating the effects of various parameters (such as the ratio of the cooling time to the free-fall time, $t_{\rm cool}/t_{\rm ff}$) 
on the evolution of over-dense blobs.

The entropy parameters and density boundary conditions for the model ICMs used in our study are stated in Table~\ref{profile_table}. The 
entropy parameters span a range covered by cool clusters in the ACCEPT sample (\citealt{accept09}).  The minimum value of the parameter 
$\tr$, which plays a very important role in governing the fate of an over-dense blob in the ICM, is also shown.  

\begin{table}
\caption{Entropy parameters and boundary density for our ICM profiles ($c = 3.3, \: M_{200} = 5.24 \times 10^{14} M_{\odot}$)}
\resizebox{0.48 \textwidth}{!}{%
\begin{tabular}{cccccc}
\hline
\hline
$K_{0} \: (\rm keV cm^{2})$ & $K_{100} \: (\rm keV cm^{2})$ & $\alpha$ & $n (r_{200}) \: (\rm cm^{-3})$  & min($t_{\rm cool}/t_{\rm ff}$) & $\beta$ parameter$^\ddag$ \\ 
\hline
1 & 115 & 1.1 & $3 \times 10^{-4}$ & 2.1 & 0.011, 0.021, 0.03 \\ 
3 & 110 & 1 & $3 \times 10^{-4}$ & 5.7 & 0.016, 0.027, 0.037 \\ 
5 & 120 & 1.15 & $3 \times 10^{-4}$ & 6.7 & 0.017, 0.027, 0.038 \\ 
$8^\dag$ & $80 $ & $1.1 $ & $1 \times 10^{-4} $ & 7.8 & 0.017, 0.025, 0.031\\  
$12.4$ & 75 & 1.3 & $3 \times 10^{-4}$ & 8.3 & 0.014, 0.025, 0.028 \\
$14.5$ & $82$ & $1.47$ & $3 \times 10^{-4}$ & 9 & 0.019, 0.026, 0.03 \\ 
$37.9$ & $117.9$ & $1.11$ & $1 \times 10^{-4}$ & 30.6 & ... \\  
\hline
\end{tabular}} \\
{\textbf{Notes}} \\
{$^\dag$}{The fiducial profile.} \\
{$^\ddag$}{The $\beta$ parameter (c.f. Eq. \ref{eq:delDotSoker_sph}) determined for blob-sizes of 10, 50, 100 pc to form cold gas at radii 
within which $\tr$ falls below 10. All our runs use the average $\beta$ value of 0.0244.}\\
\label{profile_table}
\end{table}

We use the parameters included in Table \ref{profile_table} for our models with spherical compression. However, as we show in 
section \ref{sec:tr_cart}, the nature of the evolution of the blob over-density changes if the ratio $\tr$ is below a critical value, for 
both Cartesian and spherical geometries. This critical value for Cartesian setups is 1, typically not reached by even the coolest cluster 
cores (see Table \ref{profile_table}). Therefore, in order to elucidate the physics of the evolution of over-density in a Cartesian setup 
we {\em artificially} increase the cooling function $\Lambda$ for all Cartesian runs by a factor of 10; this way we can achieve 
$\tr \lesssim 1$ required to study the runaway cooling of over-dense blobs in a plane-parallel atmosphere.

\section{Local thermal instability in a globally stable  ICM}

In the ICM the rate of energy loss due to free-free/bound-free/bound-bound radiation can be modeled as 
\begin{equation}
\label{eq:n2lambda}
P_{\rm loss} = \frac{dE}{dtdV} = - n_{e} n_{i} \Lambda(T) \: ,
\end{equation}
where $n_{e}$ and $n_{i}$ represent the electron and ion number densities and $\Lambda (T)$ is the  cooling function. 
We use the cooling function from \citealt{tozzi+norman+2001} based on  \citealt{sutherland+dopita+93}, as adapted 
by \citealt{sharma10} (solid line in their Fig. 1). In case of thermal bremsstrahlung (which dominates above $10^7$ K), the cooling 
function $\Lambda (T) \propto T^{1/2}$ and hence $P_{\rm loss} \propto n^{2} T^{1/2}$. Due to the observational lack of cooling flows, 
and given other hints of global thermal equilibrium in the ICM, we use the model of \citet{sharma12} in which  heating ({\it H}) balances 
average cooling at every radius in cluster cores,
\begin{equation}
\label{eq:H=<C>}
H = <n_{e} n_{i} \Lambda (T)>.
\end{equation}
While this clearly is an idealization, since AGN heating is expected to be intermittent with large spatial and temporal fluctuations 
around thermal equilibrium, jet simulations agree well with the simulations based on the idealized model (Eq. \ref{eq:H=<C>}; for 
comparison of jet simulations with the idealized model see \citealt{gaspari12,li14}). More importantly for this paper, our heating term 
is analytically tractable and captures the basic thermal state of the ICM.

An important parameter, which governs the evolution of linear over-dense blobs, is the ratio of the ICM cooling time $t_{\rm cool}$ 
(more precisely, the thermal instability time),
\begin{equation}
t_{\rm cool} = \frac{3}{2} \frac{ n_{0} \: k_{B} \: T_{0}}{n_{e,0} \: n_{i,0} \: \Lambda (T_{0})}
\end{equation}
 and the free fall time $t_{\rm ff}$, 
\begin{equation}
t_{\rm ff} = \left( \frac{2 \: r_{0}}{g_{0}} \right )^{1/2} 
\end{equation}
in the background ICM at the location of the blob (as highlighted by \citealt{mccourt12,sharma12}), where the quantities subscripted with `0' stand 
for their unperturbed background values at the radius under consideration. The $t_{\rm{cool}}/t_{\rm{ff}}$ profile for the fiducial cluster 
is included in Figure~\ref{fig:fiducial}.

\subsection{Evolution of a spherical blob}
We base our study on the phenomenological model of PS05 and \citealt{ps10}, who propose that over-dense blobs of gas in the ICM 
cool faster than their surroundings, become heavier and sink to the center to feed the AGN. Both these papers consider the evolution 
of blobs in a background cooling flow with some simplifications. 
In this paper we make the more realistic assumption that the core is in rough 
thermal balance without any equilibrium flow. This implicitly assumes that heating keeps up with otherwise catastrophic cooling. 

We derive the conditions in which the over-dense blobs cool to the stable atomic phase ($T < 10^{4} K$),  leading to a multiphase core. 
The conditions for the formation of the cold phase depend on the various parameters of both the blob and the ICM. We show that for a 
large $\tr$, cold gas condensation requires a finite over-density. For a sufficiently small $\tr$, however, cold gas condenses out of the 
ICM for even a tiny over-density. Moreover, we extend the PS05 formalism to account for spherical compression that makes it easier 
for cold gas to condense out in a spherical geometry as compared to a plane-parallel atmosphere.

Linear thermal instability analysis has been done in the {\it Appendix} section of \citealt{sharma10} (the original reference is 
\citealt{field65}) in the limit $t_{\rm cool} \gg t_{\rm sound}$ (the sound crossing time for the modes). In such a scenario, 
neglecting gravity, we have isobaric conditions such that the perturbation always remains in pressure equilibrium with its 
surroundings and the linear growth rate $\sigma$ of the instability is (in absence of thermal conduction),
\begin{equation}
\label{growthRate}
\sigma = -i \: \omega = \frac{- d \: \ln(\Lambda / T^{2})}{d \: \ln T} \frac{1}{\gamma  \: t_{\rm cool} } \: .
\end{equation}
As done in PS05, we consider an over-dense spherical blob of radius $a$, whose parameters will be represented with 
primed quantities ($n'\: , \: T'$), having a density contrast (over-density) with respect to the ICM (unprimed quantities represent 
the ambient ICM):
\begin{equation}
\label{eq:deltaDef}
\delta \equiv \frac{n' - n}{n}	\: .
\end{equation}
The blob is assumed to remain in pressure equilibrium at all times (i.e. $P' \equiv P$) with the ICM in the isobaric limit of 
$t_{\rm cool} \gg t_{\rm sound}$ (the blob evolves isochorically in the opposite limit, which may happen for a short time when 
the blob temperature is close to the peak of the cooling function; e.g., see \citealt{burkert00}; the isobaric assumption should 
not affect our results qualitatively), and thus, 
\begin{equation}
\label{eq:Tblob}
T' = \frac{T}{1 + \delta} \: .
\end{equation}
In a one-dimensional analysis, we re-write from PS05 the basic equations governing the blob evolution:
\begin{equation}
\label{eq:Soker1}
\frac{dr}{dt} = v \: ,
\end{equation}
\begin{equation}
\label{eq:Soker2_1}
\frac{dv}{dt} = -g \: \frac{\delta}{1 + \delta} -  {\rm sign}(v) \: \frac{3 \: C}{8 \: a} \: \frac{v^{2}}{1 + \delta} \: ,
\end{equation}
where $r$ is the radial coordinate of the blob, $v$ is the blob's radial velocity, $a$ is the blob radius, {\it C} is the dimensionless 
drag coefficient, which for most part of the paper has been taken as 0.75 (see, e.g., \citealt{chu01}), and ${\rm sign}(v)$ ensures that 
the drag force always points against the velocity. The energy equation for the over-dense blob can be written, assuming global 
thermal balance with a heating term (Eq. \ref{eq:H=<C>}),\footnote{Here we are assuming that the background ICM profile is 
not changing with time. In reality, the background profile changes because cold gas condenses out of the hot phase and falls in, 
reducing the density of the remaining hot gas. In this paper our focus is on whether cold gas can condense out in global thermal 
equilibrium for a given ICM profile, rather than on the effect of condensation on the background profile.}
\begin{equation}
\label{eq:energyEq}
\frac{P}{\gamma - 1} \: \frac{d}{dt} \: \ln \: \left [ \frac{P}{(n')^{\gamma}} \right ] = -n'_{e} \: n'_{i} \: \Lambda (T') + n_{e} \: n_{i} \: \Lambda (T)	\:\: .
\end{equation}
The primed term in the RHS of Eq.~\ref{eq:energyEq} is the cooling of the blob and the unprimed term represent the volume 
averaged heating of the ICM (and the blob). Following \citealt{ps10} and substituting $d/dt$ by $vd/dr$ for the background 
quantities, this can be recast to obtain an equation for the over-density of the blob (assuming $\gamma=5/3$):
\begin{equation}
\label{eq:delDotSoker}
\frac{d\delta}{dt}   = \frac{2 \: (1+\delta)}{5\:n\:k_{B}\:T} \: [ n'_{e} \: n'_{i} \: \Lambda (T') - n_{e} \: n_{i} \: \Lambda (T)] + \\ \frac{(1+\delta)}{g} \: v \: N^{2},
\end{equation}
where {\it $N^{2}$} is the square of the Brunt $\rm V\ddot{a}is\ddot{a}l\ddot{a}$ frequency and represents the linear response of 
an over-dense blob in a stably-stratified ICM,
\begin{equation}
\label{eq:BVOscillation}
N^{2} = \frac{g}{\gamma} \: \frac{d}{dr} \: \ln \left ( \frac{T}{n^{\gamma - 1}} \right ).
\end{equation}
In absence of cooling and heating we obtain stable Brunt $\rm V\ddot{a}is\ddot{a}l\ddot{a}$ oscillations. These oscillations are 
damped in time because of the drag term in Eq.~\ref{eq:Soker2_1}.  Although Eq.~\ref{eq:delDotSoker} is identical to Eq. 12 in 
\citet{ps10}, the interpretations are slightly different because  \citet{ps10} assume cooling of the background ICM but ignore the 
background inflow, whereas we assume global thermal balance with no net inflow of the hot gas.

The cooling and heating terms on the RHS of Eq.~\ref{eq:delDotSoker} in the linear ($\delta \ll 1$) isobaric ($n^\prime T^\prime = nT$) 
regime can be reduced to,
$$
\frac{2(1+\delta)}{5 n k_B T} n_e n_i T^2 \left[ \frac{\Lambda(T^\prime)}{T^{\prime 2}} - \frac{\Lambda(T)}{T^{2}} \right] =  \frac{1}{\gamma t_{\rm cool}} \left ( 2 - \frac{d \ln \Lambda}{d \ln T}  \right)  \delta,
$$
which when plugged in Eq.~\ref{eq:delDotSoker} gives the correct growth rate for the thermal instability in the isobaric 
regime (Eq.~\ref{growthRate}). Linearized versions of Eqs. \ref{eq:Soker2_1} \& \ref{eq:delDotSoker}, using above, give 
the expression for thermal instability (more precisely, over-stability) in a stably stratified atmosphere (e.g., see \citealt{binney09}),
\begin{equation}
\label{eq:LTI}
\frac{d^2 \delta}{dt^2} + N^2 \delta = \frac{1}{t_{\rm TI}} \frac{d \delta}{dt},
\end{equation} 
where $t_{\rm TI}=\gamma t_{\rm cool}/(2 - d\ln \Lambda/d\ln T)$.\footnote{In a local linear stability analysis the blob corresponds to a mode 
with the wavenumber perpendicular to the direction of gravity. Buoyant response is even weaker for modes with a finite wavenumber along 
the direction of gravity.}  For a heating rate per unit volume proportional to 
density (e.g., as is the case for photoelectric heating), the heating term in Eq.~\ref{eq:energyEq} is $(1+\delta)n_e n_i \Lambda(T)$ 
and the corresponding linear analysis gives $t_{\rm TI} =\gamma t_{\rm cool}/(1 - d\ln \Lambda/d\ln T)$, the correct analytic 
result (e.g., see \citealt{mccourt12}). Thus, strictly speaking, the cold gas over-density grows at the thermal instability timescale 
($t_{\rm TI}$) and not at the cooling time ($t_{\rm cool}$); this is indeed verified by the simulations of \citealt{mccourt12}. 
We use $t_{\rm cool}$ instead of $t_{\rm TI}$ throughout the paper because $t_{\rm TI} = (10/9) t_{\rm cool} \approx t_{\rm cool}$ 
for the case of a constant heating rate per unit volume. Local thermal instability as the source of extended cold gas in cluster cores, 
with a constant heating rate per unit volume, is supported by observations (Fig. 11 in \citealt{mccourt12}). Similar scaling of the 
heating rate with the local density is supported by the simulations of turbulent heating/mixing in equilibrium with 
cooling (\citealt{banerjee14}).

During its evolution, the blob mass should be conserved. For the spherical blob of mass {\it M}, we have,
\begin{equation}
\label{blobMass}
M = \frac{4 \pi}{3} \: a^{3} \: n \: (1\: + \: \delta) \mu m_{p} = \frac{4 \pi}{3} \: {a_{0}}^{3} \: n_{0} \: (1\: + \: \delta_{0}) \mu m_{p},
\end{equation}
where quantities with subscript `0' represent their initial/background values. 

 Eqs.~\ref{eq:Soker1}, ~\ref{eq:Soker2_1}, ~\ref{eq:delDotSoker} and ~\ref{blobMass} represent a system of 4 equations 
in 4 variables ({\it ${r[t],\: v[t],\: \delta[t],\: a[t]}$}) that can be solved numerically as an initial value problem to study the evolution of 
blobs in the ICM. We specify the initial values $r_0$, $\delta_0$ and $a_0$; initial velocity is chosen to be zero because cold gas 
is condensing out of the gas in hydrostatic equilibrium. In absence of the drag term, Eqs. (\ref{eq:Soker1}), (\ref{eq:Soker2_1})
and (\ref{eq:delDotSoker}) do not depend on the blob-size, and blobs of different size evolve in a similar fashion. Drag slows 
down smaller blobs and give them more time to cool before they can fall in. Drag also damps stable oscillations.

\section{Results}
In this section we present a phenomenological model to include the effects of geometrical compression in a  spherical ICM. Later, 
we present the results on the evolution of over-dense blobs in both Cartesian and spherical geometries. The evolution and 
saturation of the over-density ($\delta$) of the blob is a sensitive function of the background $\tr$, the initial over density 
$\delta_{0}$ and the spherical compression term; there is much weaker dependence on the initial blob radius {\it $a_{0}$}, 
the drag coefficient and entropy stratification (because it is weak in the ICM, as we discuss later). We construct models in 
which runaway cooling occurs starting with tiny amplitude of perturbations only if the 
background $\tr \lesssim 1,~10$ in Cartesian and spherical atmospheres, respectively. If $\tr$ is larger than the critical 
value for runaway at tiny amplitudes, a finite amplitude of the over-density is required for runaway.

\citet{mccourt12} and \citet{sharma12} have carried out idealized simulations of hot atmospheres in thermal and hydrostatic 
equilibrium, using Cartesian and spherical geometries, respectively. Somewhat surprisingly, they find that it is much easier 
for cold gas to condense out of the hot phase in a spherical geometry. This, they attribute to spherical compression that an 
over-dense blob undergoes as it moves toward the centre in a spherical geometry. More quantitatively, they find that cold 
gas can condense out from small initial perturbations if $t_{\rm cool}/t_{\rm ff} \lesssim 1$ in Cartesian geometry and if 
$t_{\rm cool}/t_{\rm ff} \lesssim 10$ for a spherical setup. 

PS05 do not consider any influence of geometry in their analysis, and therefore their results do not depend on whether 
the blob is falling in a plane-parallel atmosphere or a spherical one. Unlike in a Cartesian atmosphere, in spherical geometry 
the gravitational forces at the diametrically opposite ends of a spherical blob (at the same height) are not parallel. The radial
component of gravity pointing toward cluster center will compress the blob and make it further over-dense, in addition to the 
effects already accounted for in Eq.~\ref{eq:delDotSoker}. Thus, an over-dense blob in a spherical geometry is compressed 
more than the one in a plane-parallel atmosphere, and forms cold gas more easily. 
	
\subsection{Modeling geometric compression}
\label{sec:sph_comp}

In this section, we present a simple phenomenological prescription that allows us to parametrically model the blob's 
compression in a spherical geometry. Considering only geometrical 
compression, the transverse cross-section of the blob (which subtends a constant solid angle at the centre as it falls in) 
decreases as it falls in toward the centre, and the blob over-density increases such that $(1+\delta) r^2=$ constant ($r$ is 
the radial coordinate of the centre of the blob). The corresponding expressions for cylindrical and Cartesian geometries 
are $(1+\delta) R=$ constant ($R$ is the cylindrical radius) and $(1+\delta)=$ constant. This implies that there is no 
geometrical compression in the Cartesian geometry. The compression in cylindrical geometry is smaller than in  
a spherical atmosphere because over-dense blobs are compressed to a line and not a point as they fall in. Numerical 
simulations in cylindrical geometry indeed show that the threshold $t_{\rm cool}/t_{\rm ff}$ 
for the production of multiphase gas with tiny initial amplitude lies in between the results from Cartesian and spherical setups 
(M. McCourt, private communication). 
Our model fine-tuned for spherical profiles shows that the critical value of $\tr$ for condensation of cold gas in a cylindrical geometry, starting 
from tiny amplitudes, 
ranges from 2.5 to 5 (as compared to the Cartesian and spherical cases, this value is somewhat sensitive to the background entropy profile; 
the analogous values for Cartesian and spherical atmospheres are 1 and 10, respectively; c. f. Figs. \ref{trcrit_cart}, \ref{trcrit_sph}).

\begin{figure}
  \includegraphics[scale=0.34]{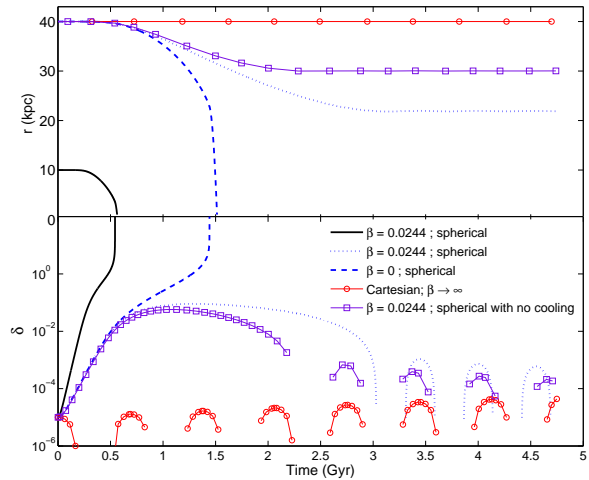}
  \centering
  \caption{The location of the over-dense blob ({\em top panel}) and the over-density ({\em bottom panel}) as a function 
  of time for fiducial models with and without spherical compression, and without the modulation of the spherical 
  compression term (i.e., $\beta=0$). The $\beta$ parameter (Eq. \ref{eq:delDotSoker_sph}) for some runs is 
  chosen ($\beta=0.0244$) such that small amplitude perturbations outside $\tr =10$ do not condense out. The initial blob-size 
  is $a_0=50$ pc; results are only weakly dependent on $a_0$. The blob evolution is also shown for the spherical compression 
  model without cooling.
  }\label{vstime}
\end{figure}

Therefore, in order to model spherical compression, we should add a term like
$$
\left ( \frac{d \delta}{dt} \right)_{\rm sph} = -2 \frac{(1+\delta)}{r} \frac{dr}{dt} = -2 \frac{ (1+\delta) v}{r},
$$
to the right hand side of Eq.~\ref{eq:delDotSoker}. This term is negligible if the entropy scale height ($1/[d\ln K/dr]$) is much smaller than $r$; 
or in other words, if the atmosphere is strongly stratified, as we show shortly. Hence, in spherical geometry,
\begin{eqnarray}
\label{eq:delDotSoker_sph}
\nonumber
\frac{d\delta}{dt}   &=& \frac{2 \: (1+\delta)}{5\:n\:k_{B}\:T} \: [ n'_{e} \: n'_{i} \: \Lambda (T') - n_{e} \: n_{i} \: \Lambda (T)] \\
&+& \frac{(1+\delta)}{g} \: v \: N^{2} - 2 \frac{ (1+\delta) v}{r} e^{(-\beta M_{\rm encr}/M)},
\end{eqnarray}
where we have added a phenomenological modulation term (similar in spirit to the widely-used mixing-length models) to the 
compression term, which suppresses geometrical compression when the mass encountered by the blob in the ICM becomes 
comparable to the blob's mass. This term is motivated by the fact that hydrodynamic effects, such as distortion due to ram 
pressure and the loss of sphericity of the blob, are expected to become dominant over geometrical compression, once 
the mass encountered is of order the blob mass. The $\beta$ parameter will be chosen so that the model gives results 
consistent with \citet{sharma12}; namely, the over-dense blobs with tiny amplitudes run away to the stable temperature if 
and only if $t_{\rm cool}/t_{\rm ff} \lesssim 10$. The mass encountered by the blob is given by integrating
\begin{equation}
\label{Mencounter}
\frac{dM_{\rm encr}}{dt} \: = \: \pi \: a^{2} (t) \: \rho (r(t)) \: \big\lvert v (t) \big\rvert \: ,
\end{equation}
where $\big\lvert v (t) \big\rvert$ ensures that the mass encountered by the blob increases monotonically; 
this is required because hydrodynamic distortion happens irrespective of the direction of blob motion. 
Thus, in spherical geometry we need to solve 5 equations, namely Eqs. ~\ref{eq:Soker1}, \ref{eq:Soker2_1},
~\ref{blobMass},~\ref{eq:delDotSoker_sph}, and \ref{Mencounter} for 5 unknowns (the additional unknown 
compared to PS05 being $M_{\rm encr}$).

The spherical compression term (the last term in Eq. \ref{eq:delDotSoker_sph}) can be linearized as $-2 v/r$. The 
linearized equation governing the evolution of $\delta$, including spherical compression, becomes
\be
\label{eq:linear_sph}
\frac{d^2 \delta}{dt^2} + \left ( N^2 - \frac{2 g}{r} \right ) \delta = \frac{1}{t_{\rm TI}} \frac{d \delta}{dt},
\ee
where we have used the linearized equation of motion $dv/dt = -g \delta$. From Eq. \ref{eq:BVOscillation}, $N^2 = g/(\gamma H)$,
where $H=1/(d \ln K/d\ln r)$ is the entropy scale height. It is clear from Eq. \ref{eq:linear_sph} that the spherical compression term
leads to linear instability if $H>r/(2\gamma)$, even in the absence of thermal instability. As expected, the local plane-parallel approximation, 
with a negligible effect of spherical compression, holds in the limit $H \ll r$. The entropy scale height of the ICM is 
comparable to $r$, and hence the amplitude of over-dense blobs grows linearly because of spherical compression; the amplitude 
saturates nonlinearly when the blob encounters its own mass in the ICM and the spherical compression term in Eq. 
\ref{eq:delDotSoker_sph} is suppressed (see the line with square markers in Fig. \ref{vstime} which shows the evolution with 
spherical compression in the absence of cooling). When $\tr \gtrsim 10$ over-density never becomes large or becomes large only when 
blob falls to the very center. However, with $\tr \lesssim 10$ local thermal instability can lead to extended cold gas.

The $t_{\rm cool}/t_{\rm ff}$ profiles for cool-core clusters have a characteristic shape with a minimum in $\tr$ at 
$\sim 10$ kpc (see Fig. \ref{fig:fiducial}). There are two points where $\tr=10$, on either side of the minimum. We find 
out the value of $\beta$ for each cluster such that the blob released at the outer point where $\tr \approx 10$ (lets call it 
$r_{10}$) would just form cold gas on reaching 1 kpc.\footnote{The choice of 1 kpc as the threshold for defining {\em extended} 
cold gas is observationally motivated as the observed multiphase gas is spread over scales of $\sim$10 kpc (\citealt{mcdonald10}).} 
For such a value of $\beta$, no cold gas forms when $r_0 \: > \: r_{10}$, i.e., the region where $\tr \:> \: 10$. We chose the 
outer $r_{10}$ to determine $\beta$ because the outer regions with $\tr \lesssim 10$ are most likely to form {\em extended} 
cold gas. For starting radii well within the bottom of the $\tr$ `cup' and for small over-densities the infall time is shorter than 
the cooling time, and runaway happens very close to the center (at $r < 1$ kpc).

For the analytic model to be useful, the value of $\beta$ should not vary for different clusters and blob sizes. We list the $\beta$ 
values for various ICM profiles and blob parameters in Table \ref{profile_table}. Note that the smaller blobs require a smaller $\beta$; 
i.e., higher compression, to form extended cold gas. This is because a smaller blob encounters its own mass in the ICM faster 
than a bigger one. For all our runs we use the average of all $\beta$s listed in Table \ref{profile_table}; i.e., $\beta=0.0244$. 
As we show, our results are not very sensitive to the exact value of $\beta$.


Figure \ref{vstime} shows the evolution of 50 pc blobs with a small initial over-density ($\delta_0=10^{-5}$), 
released from initial radii of 10 and 40 kpc in the fiducial ICM with spherical compression. The evolution of 
the blobs is qualitatively similar to those in PS05, except that cold gas condenses for a longer cooling time 
because of spherical compression. If the cooling time is long, the over-density grows (because of spherical compression) 
but saturates with a small 
amplitude, and the blob undergoes stable oscillations (dotted lines); the blob falls in from its initial position 
by a substantial distance because it becomes further over-dense due to the spherical compression term. 
The blob released at the location (10 kpc) with a smaller $\tr~(<10;$
see Fig. \ref{fig:fiducial}) increases its over-density to a large value ($\gtrsim 10^4$) and falls in toward the 
center at an accelerated rate (solid line). 

\begin{figure}
  \includegraphics[scale=0.35]{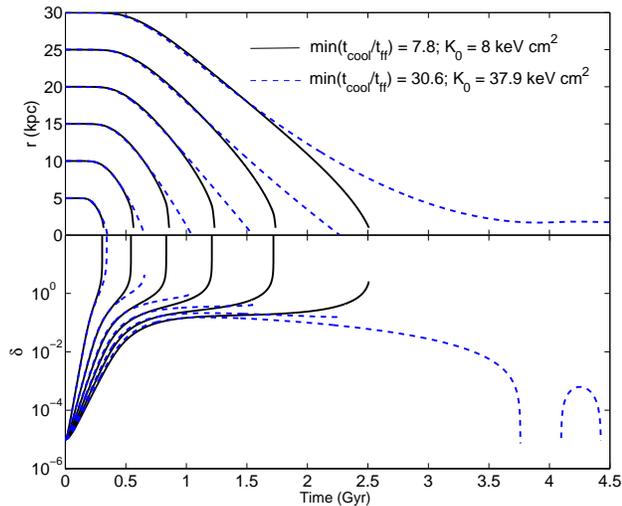}
  \centering
  \caption{The evolution of the blob location (upper panel) and the blob over-density (lower panel) as a function of time for 
  the blobs released at 5, 10, 15, 20, 25, 30 kpc: solid lines correspond to the fiducial model with $K_0=8$ keV cm$^2$; 
  dashed lines correspond to the model with $K_0=37.9$ keV cm$^2$. The lower panel shows that thermal runaway for higher 
  $K_0$ occurs only for a release radius of 5 kpc; moreover, runaway occurs as the blob reaches the very center ($\ll$ 1 kpc). 
  Runaway cooling occurs for all release radii except $r_0=30$ kpc (where $\tr > 10$) for the fiducial ($K_0=8$ keV cm$^2$) model. 
  The blob-size is 50 pc and we include spherical compression with $\beta=0.0244$ (see Eq. \ref{eq:delDotSoker_sph}).}\label{vstime_hiK}
\end{figure}

In Figure \ref{vstime} the  dotted and dashed lines correspond to the blobs released at the same radius (40 kpc) 
but with and without the $\beta$ modulation term (Eq. \ref{eq:delDotSoker_sph}). The $\tr$ ratio is 13.4 at the 
release radius of 40 kpc; the $\beta$ parameter is chosen such that the blobs released outside of the $\tr=10$ region 
do not run away. For $\beta=0$, spherical compression acts even at smaller radii and makes the blob cool in a runaway 
fashion even if $\tr > 10$, but $\beta=0.0244$ recovers the result from numerical simulations; i.e., runaway cooling for small initial 
over-densities happens only if $\tr < 10$. Therefore, the inclusion of the modulation term in Eq.~\ref{eq:delDotSoker_sph}
is necessary to get a quantitative match with numerical simulations. The value of $\beta=0.0244$ is the mean of $\beta$s 
that were obtained by requiring cold gas ($\delta \gtrsim 10^4$) to form only if $\tr < 10$ for the profiles in Table 
\ref{profile_table} (for blob-sizes of 10, 50, 100 pc); the $\beta$ range is narrow, from 0.021 to 0.028 for a blob-size of 
50 pc, and does not vary systematically for a given blob-size. 

Figure \ref{vstime_hiK} shows the blob location and the over-density as a function of time, with different initial radii, 
for the fiducial profile and for the profile with $\tr > 10$ everywhere. As expected, with $\beta$ adjusted such that extended 
cold gas is produced only if $\tr < 10$, all initial radii for the fiducial profile except for 30 kpc (at which the background 
$\tr>10$) lead to the formation of multiphase gas. The blob released at 5 kpc cools to the stable temperature very close to the center. 
Even in the most thermally unstable cases, multiphase gas appears only at $r \lesssim 5$ kpc. For the high entropy model shown 
in Figure \ref{vstime_hiK} none of the blobs released at radii $> 5$ kpc produce cold gas, 
even when they fall in toward the center; the cooling time is longer than the inflow time. 
We can relate the cold gas condensing out at large radii ($> 1$ kpc) and at small radii ($<1$ kpc) to the observational appearance 
of extended atomic filaments and centrally concentrated cold gas, respectively, as observed by \citealt{mcdonald10}. Thus, 
the results of our phenomenological model are consistent with the observational and computational results, which show that 
extended cold gas condenses out only if $\tr \lesssim 10$. Cold gas condenses out much farther out if the core density is high 
and $\tr \lesssim 10$ close to 100 kpc. 

\subsection{Extended cold filaments with tiny over-density: the critical $\tr$}

There are three ways in which an over-dense blob evolves in the ICM: it can saturate at low amplitude ($\delta < 1$) away from 
the cluster center or becomes dense only at small radii ($<1$ kpc) as it falls in; it can cool to a large over-density far from the center 
($>1$ kpc) and then fall in; it can become under-dense and overheated such that $\delta \rightarrow 1$ as it moves out.  We expect 
thermal runaway to happen because the ICM is locally thermally unstable. Only below $10^4$ K, below which the cooling function 
decreases rapidly ($\propto T^6$), is the blob expected to stop cooling further. In isobaric conditions the blob temperature $T'$ and 
the over-density ($\delta$) have an inverse proportionality, the cold gas can reach a maximum over density $\delta \gtrsim 100$ 
(positive runaway)\footnote{Once the over-dense blob reaches $\delta \sim 10$, it cools very rapidly to the stable temperature 
(see Fig. \ref{vstime_hiK}). Therefore, an over-density of 100 is reached at almost the same radius at which the blob cools to 
$T^\prime \approx 10^4$ K. Moreover, a different $\delta$ corresponds to the temperature of the stable phase; e.g., $\delta$ corresponding 
to the stable phase in a 1 keV group is $\approx 10^3$ and in a $10^6$ K galactic halo is $\approx 100$.} due to cooling. On 
the other hand, if the blob starts becoming under-dense with time, it rushes away from the ICM 
center due to its lower density, and in this case $\delta \rightarrow -1$, implying a negligibly small density of the blob. In both these 
extreme cases of runaway, the blob either forms hot or very cold gas compared to the background ICM. These runaway cases are 
required for the formation of {\em extended} multiphase gas in the ICM. We define a `multiphase flag' (mp) to assess the formation of 
{\em extended} multiphase gas in our models as follows:
\begin{equation}
\label{eq:mpFlag}
\rm mp =
\begin{cases}
+1 & \text{;  positive runaway, $\delta \gtrsim 100$ outside $r=1$ kpc} \\
0 & \text{;   no runaway outside 1 kpc } \\
    & \text{ and/or  stable oscillations} \\
-1 & \text{;  negative runaway,  $\delta \rightarrow -1$}
\end{cases}
\end{equation}
We focus on positive runaways in which there is formation of cold gas ($T' \: < \: 10^{4} \: K$), which sinks and feeds the central 
black hole required for the feedback cycle to close, as discussed earlier. 

As already mentioned, \citet{mccourt12} and \citet{sharma12}, in their Cartesian and spherical simulations of locally unstable 
hydrostatic gas in global thermal balance, find that runaway cooling of even slightly over-dense blobs occurs if the ratio of the 
cooling time to the free-fall time ($\tr$) is smaller than a critical value. This critical value, $(\tr)_{\rm crit}$, is dependent on the 
geometry of the gravitational potential in a weakly stratified atmosphere such as the ICM. Namely, multiphase gas condenses 
out of the ICM if $\tr \lesssim 10$ in spherical potential and if $\tr \lesssim 1$ in Cartesian geometry.


\subsubsection{Critical $\tr$ in a plane-parallel atmosphere}
\label{sec:tr_cart}
We use the same set of equations as PS05 to study the evolution of over-dense blobs in a plane-parallel atmosphere. PS05 
emphasized that the ICM required nonlinear density perturbations in order for multiphase gas to condense out of the ICM. 
However, they missed the significance of spherical compression brought to the fore by the idealized simulations of 
\citet{mccourt12,sharma12}, and included in this paper phenomenologically in Eq.~\ref{eq:delDotSoker_sph}. As shown by 
numerical simulations, and as we show later, this new compression term can lead to the condensation of cold gas from arbitrarily 
small perturbations in cluster cores if $t_{\rm cool}/t_{\rm ff} \lesssim 10$. Additionally, and understandably, because galaxy 
cluster cores have $\tr \gtrsim 10$, PS05 overlooked that there was another regime for the local thermal instability where arbitrarily 
small perturbations can lead to the condensation of cold gas even in a plane-parallel atmosphere.  We find, in agreement with 
\citet{mccourt12}, that in a plane-parallel atmosphere cold gas condenses out starting from a small amplitude 
if $t_{\rm cool}/t_{\rm ff} \lesssim 1$.

\begin{figure*}
\includegraphics[scale=0.5]{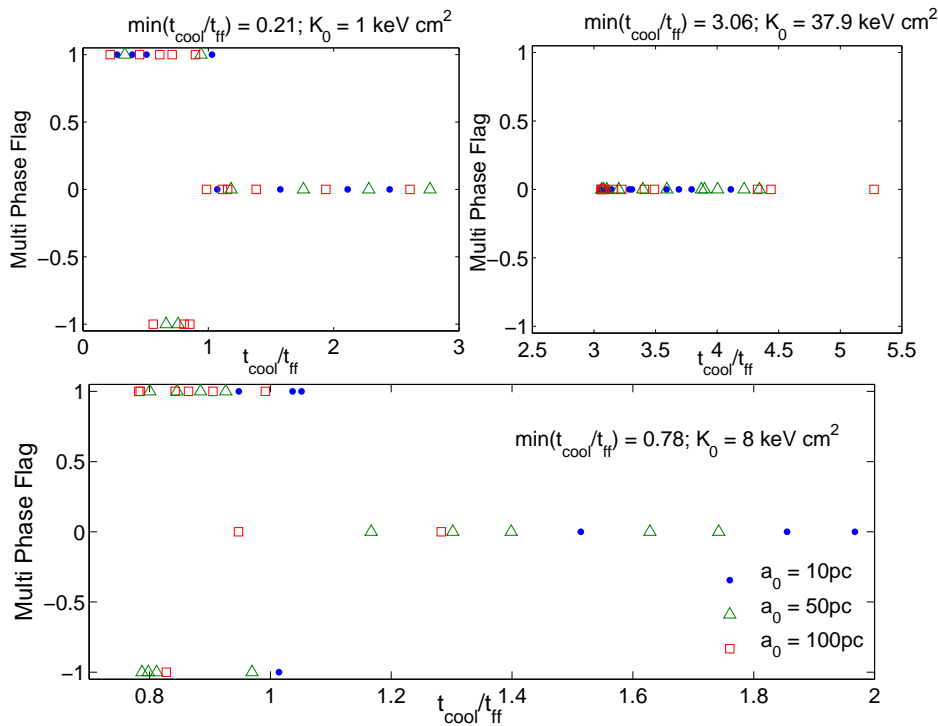}
  \caption{Multiphase flag (labelled with the initial blob size) as a function of $\tr$ at the release radius in a plane-parallel ICM for 
  three different profiles. Runaway cooling of the blobs with small initial over-densities ($\delta=10^{-5}$ is chosen) happens only 
  if the ratio $\tr \lesssim 1$; the results are not sensitively dependent on the blob-size or the drag-coefficient 
  ($C$ in Eq.~\ref{eq:delDotSoker_sph}). Note that for the same $\tr$ and blob-size we can have a positive or a negative runaway; 
  this is because the blobs with release radii on either side of min($\tr$) evolve differently (see Fig. \ref{fig:37}). Since observed clusters 
  have $\tr \gtrsim 5$, we have {\em artificially} increased the cooling function ($\Lambda[T]$) by a factor of 10 to attain 
  $\tr \lesssim 1$ in our plane-parallel runs.} 
\label{trcrit_cart}
\end{figure*}

As already mentioned, we achieve $t_{\rm cool}/t_{\rm ff} \lesssim 1$ in the ICM by {\em artificially} increasing the cooling 
function $\Lambda$ by a factor of 10 for plane-parallel models. We start with a tiny over-density ($\delta_0=10^{-5}$), and 
initialize the blobs at multiple radii $r_0$, and integrate the blob evolution equations for 5 Gyr for different 1-D ICM profiles 
given in Table \ref{profile_table}. In Figure \ref{trcrit_cart} we plot the multiphase flag as a function of $\tr$ (at the initial location 
where the blob is released) in a Cartesian potential, starting with a small over-density. It is evident  that for $\tr > 1 $, there are 
no runaways seen for small $\delta_{0}$. The critical $\tr$ in a Cartesian gravity setup using the simple model of PS05 
is  $(\tr)_{\rm crit} = 1$. The critical value of $\tr$ is fairly insensitive to the blob size, the initial over-density 
and the drag coefficient.
For $\tr \lesssim 1$ thermal instability forms multiphase gas irrespective of the blob parameters, as seen in the simulations of 
\citet{mccourt12}. 

\subsubsection{Critical $\tr$ in a spherical ICM}

\begin{figure*}
\includegraphics[scale=0.5]{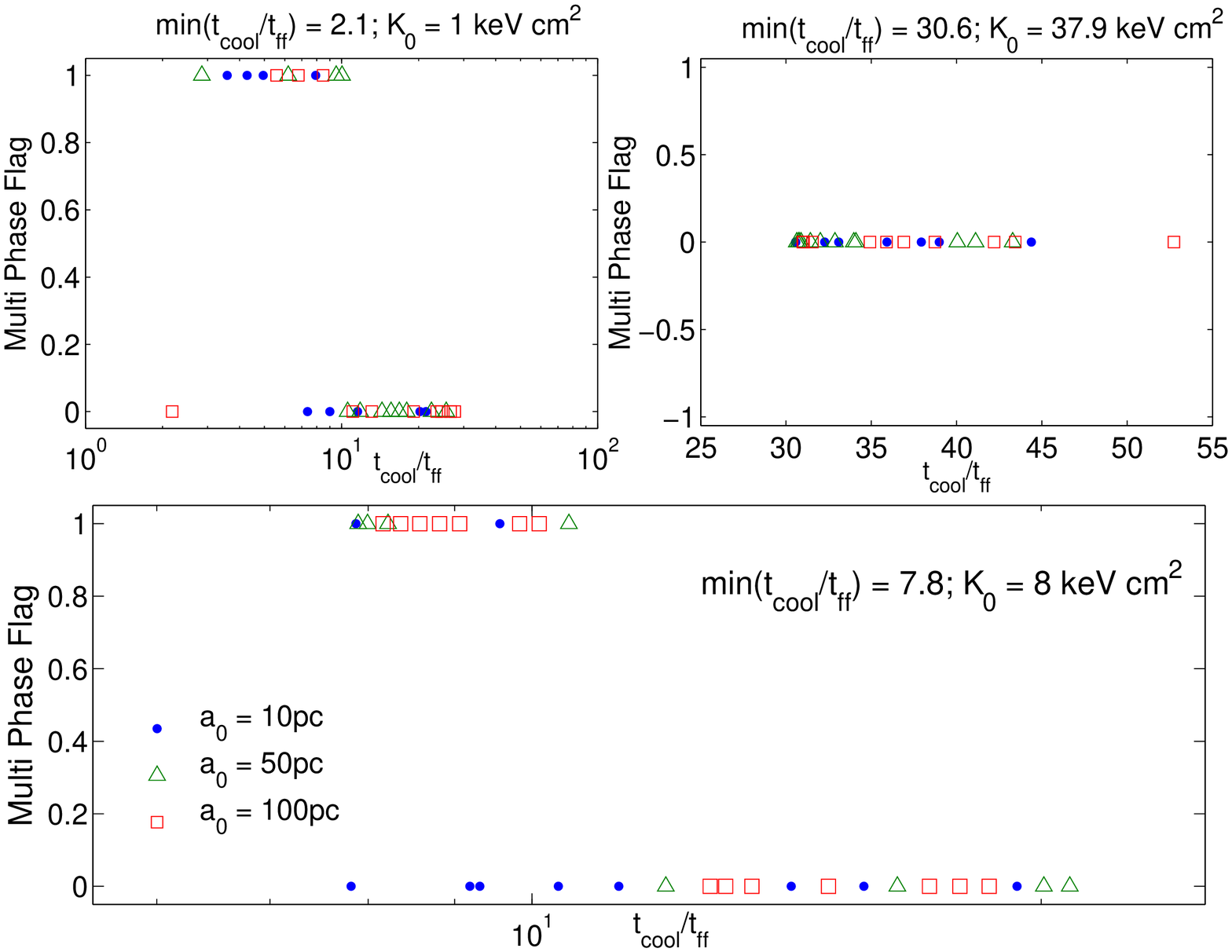}
  \caption{Multiphase flag (labelled with the initial blob size) as a function of $\tr$ in spherical gravity for $\delta_{0}=10^{-5}$. 
  Runaway cooling in spherical geometry happens for a higher $\tr$ ($\lesssim 10$); i.e., in comparison to a plane-parallel atmosphere, 
  cold gas condenses out more easily in a 
  spherical atmosphere. Also note, in comparison to the plane-parallel atmospheres (Fig. \ref{trcrit_cart}), there are no hot runaways 
  in spherical geometry because of the geometrical compression term in Eq.~\ref{eq:delDotSoker_sph}. }\label{trcrit_sph}
\end{figure*}

We now incorporate the geometrical compression model discussed earlier in section \ref{sec:sph_comp}, to study blob 
evolution in a spherical setup and determine the critical $\tr$ for runaway to occur with tiny over-densities. Using spherical 
simulations \citealt{sharma12} find the critical value for the condensation of cold gas to be $(\tr)_{\rm crit} \approx 10$. 
This critical criterion agrees with the observations of cool cluster cores which show extended cold filaments (Fig. 11 
in \citealt{mccourt12}). 

Our compression model includes a free parameter $\beta$, which we adjust to match the critical $\tr$ from simulations and 
observations, as discussed in section \ref{sec:sph_comp}. For all our spherical models we use $\beta=0.0244$, the average 
of $\beta$s obtained for various blob and ICM parameters. Clusters with $K_{0} \: \gtrsim \: 30 \: \rm keV cm^{2}$, have $\tr > 10$ 
everywhere. Hence, extended cold phase is not expected in such clusters for small amplitudes. Our models with spherical compression 
show a similar trend. Figure \ref{trcrit_sph} shows the results of our model by plotting the multiphase flag of Eq.~\ref{eq:mpFlag}
against $\tr$ for over-dense blobs starting at various positions in some of our clusters in Table \ref{profile_table}. It is evident that 
multiphase condensation takes place for arbitrarily small amplitudes only when $\tr \lesssim 10$, as found in numerical simulations 
by \citealt{sharma12}.

\subsection{Multiphase gas with long cooling times}

\begin{figure}
  \includegraphics[scale=0.3]{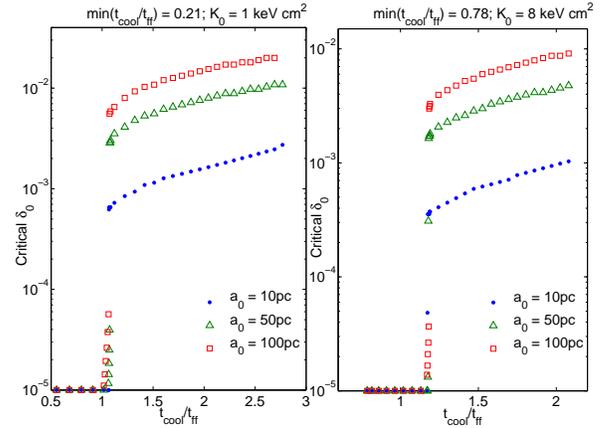}
  \centering
  \caption{The critical value of the initial over-density ($\delta_0$) required for runaway cooling as a function of $\tr$ in a 
  plane-parallel atmosphere. Runaway cooling happens only if the over-density is larger than this critical value; for a smaller 
  $\delta_0$, the blob undergoes linear oscillations.}\label{deltacrit_cart}
\end{figure}

In cases where the background $\tr$ is higher than the critical value, there is a competition between cooling and free-fall, and cooling 
to low temperatures occurs only with a finite initial over-density ($\delta_0$). This should be contrasted with the cases where 
$\tr \lesssim t_{\rm cool}/t_{\rm ff, crit}$, and cold gas condenses out even with a tiny over-density. We release stationary blobs 
at varying initial radii, $r_0$ (ranging from 10 to 100 kpc), to achieve a spectrum of $\tr > t_{\rm cool}/t_{\rm ff, crit})$ and track 
their evolution for 5 Gyr in both Cartesian and spherical potentials.

In both Cartesian and spherical atmospheres, the trends seen for the critical initial over-density ($\delta_{0,c}$) required to 
form extended cold phase outside of 1 kpc are similar. This is intuitively expected; as $\tr$ increases, cooling becomes less 
effective and the blob needs a higher density contrast to cool faster than it can fall in to form extended cold gas.
The geometrical compression term in Eq.~\ref{eq:delDotSoker_sph} increases the critical value of $\tr$, above which a finite large 
over-density is required to form cold phase. 

In Figure~\ref{deltacrit_cart}, we show the dependence of $\delta_{0, \rm c}$ on the  background $\tr$ for the fiducial and $K_0=1$ keV cm$^2$ 
 profiles using a variety of blob sizes. It is seen that in the absence of geometrical compression, smaller blobs cool more easily. 
The blob-size enters the equations via the drag term in the equation of motion (Eq. \ref{eq:Soker2_1}). A smaller blob (for a given over-density), 
or equivalently a larger drag term, implies that the blob is slowed down more as it is falling in, and therefore has a longer time to cool compared 
to a bigger blob. This explains a smaller critical $\delta_0$ for smaller blobs. 

\begin{figure}
  \includegraphics[scale=0.3]{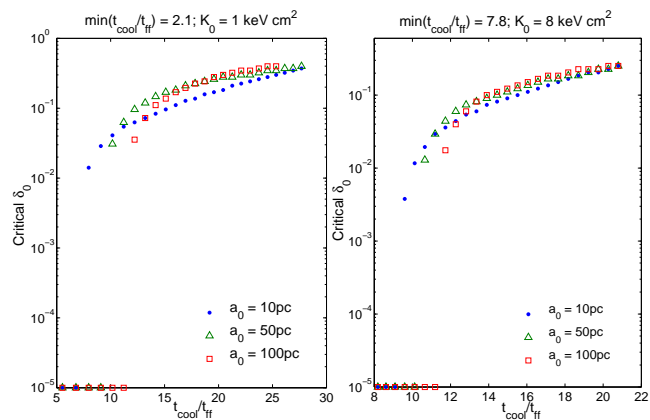}
  \centering
  \caption{Critical Initial over-density (blob-size is labelled) $\delta_{0,\rm c}$ for different $\tr$ for a spherical ICM. Notice that the 
  larger blobs form extended cold gas for a slightly higher $\tr$.}\label{deltacrit_sph}
\end{figure}

Figure~\ref{deltacrit_sph} shows the critical over-density ($\delta_{0,c}$) required for runaway cooling in a spherical ICM. We include the 
geometrical compression term with modulation due to the encountered mass (the last term in Eq. \ref{eq:delDotSoker_sph}). For larger blobs 
the compression term acts for a longer time and tends to compress it more because it takes longer to sweep up a bigger blob's own mass in 
the ICM. On the other hand, like in a plane-parallel atmosphere, the lower drag makes it fall faster. These two effects oppose each other, 
and we see that the critical $\delta_{0}$ required to form cold phase in spherical geometry is insensitive to the blob size at higher $\tr$. In 
both Cartesian and spherical atmospheres, the important point to note is the sharp fall of $\delta_{0, \rm c}$ near the critical $\tr$. It is 
evident that below the critical value of $\tr$, any arbitrary small value of $\delta_{0}$ will form cold phase. This shows the existence of 
the critical $\tr$ and the domination of thermal instability growth for a small $\tr$.

\begin{figure}
  \includegraphics[scale=0.36]{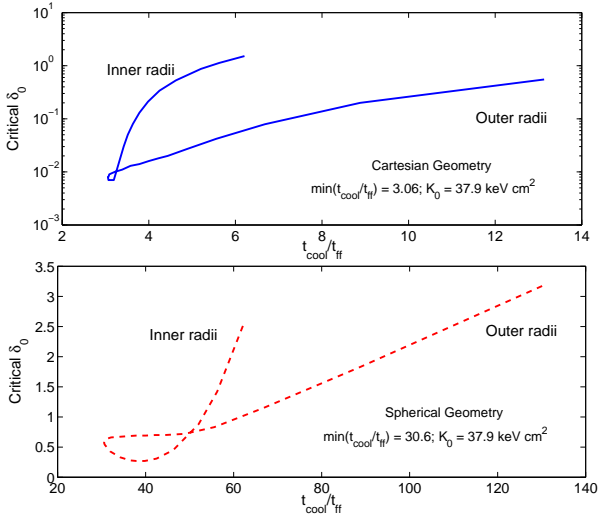}
  \centering
  \caption{The critical value of the over-density as a function of $\tr$ for the high entropy model in Cartesian (upper panel) and 
  spherical (lower panel) geometries.  For the same value of $\tr$ the blobs starting farther away from the minimum of $\tr$ require 
  a smaller over-density for runaway cooling as compared to the blobs released at smaller radii.  Note that the vertical scale is 
  logarithmic (linear) for the plane-parallel (spherical) atmosphere.}\label{fig:37}
\end{figure}

Figure \ref{fig:37} shows the critical over-density required for runaway cooling as a function of $\tr$ in spherical and Cartesian 
geometries for the highest entropy model in Table \ref{profile_table}. The critical over-density has a characteristic shape for both 
Cartesian and spherical atmospheres. Also, note that the minimum over-density required for runaway cooling does not 
correspond to the release radius with minimum $\tr$; it corresponds to a radius slightly inward of the minimum. The asymmetry 
in the response of over-dense blobs inward and outward of the minimum is likely responsible for the presence of {\em extended} 
cold filaments in some cool cluster cores (\citealt{mcdonald10}). Somewhat counter-intuitively, 
the blobs farther away from the center (with a slightly longer cooling time than at the center) are more likely to result in cold gas
because the blobs at inner radii fall in before they can cool to the stable temperatures. This leads to spatially 
extended cold gas over 10s of kpc, rather distinct from the centrally concentrated cold gas.

\section{Discussion \& Conclusions}

In this paper we have presented a phenomenological model for the evolution of over-dense blobs in the ICM. We have extended 
the model of \citet{ps05} to include the important influence of geometrical compression. A comparison of idealized simulations in 
spherical (\citealt{sharma12}) and plane-parallel (\citealt{mccourt12})  atmospheres show that it is much easier for multiphase gas 
to condense in the presence of spherical compression. We have incorporated a phenomenological spherical compression term in 
the model of PS05; this increases the over-density as the blob falls in toward the center. With a single adjustable parameter ($\beta$ 
in Eq.~\ref{eq:delDotSoker_sph}), which is analogous to the mixing length, we are able to obtain the key result of \citet{sharma12}; i.e., 
cold gas condenses out at large radii starting from tiny perturbations when the ratio of the cooling time and the local free-fall time 
($t_{\rm cool}/t_{\rm ff}$) in the ICM is $\lesssim 10$.

In the following we discuss various astrophysical implications of our results.
\begin{enumerate}
\item {\em Robustness of the phenomenological model:} Our phenomenological model, which extends PS05's treatment to account for spherical compression, agrees well with the results of numerical simulations with just a single adjustable parameter ($\beta$; see Eq. \ref{eq:delDotSoker_sph}) analogous to the mixing length. Moreover, the parameter is fairly insensitive to various parameters such as the blob-size, release radius, entropy profiles, halo mass, etc.

\begin{figure}
  \includegraphics[scale=0.35]{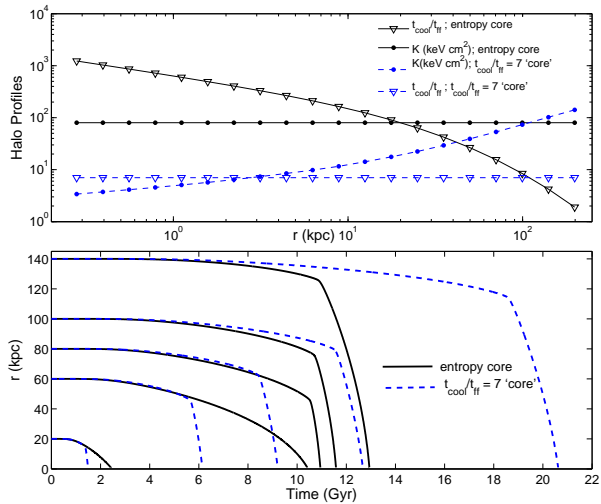}
  \centering
  \caption{The top panel shows the entropy (in keV cm$^2$) and $t_{\rm cool}/t_{\rm ff}$ profiles as a function of radius for 
  two lower mass halo ($10^{12} \msun$) models accounting for spherical compression ($\beta=0.0244$; see Eq. 
  \ref{eq:delDotSoker_sph}): one with a constant entropy core ($K=80$ keV cm$^2$) and another with a constant 
  $t_{\rm cool}/t_{\rm ff}=7$ `core'. The bottom panel shows the location of the blob (with initial radius 50 pc and initial 
  over-density $\delta=10^{-5}$) as a function 
  of time for different initial release radii for these two different profiles. While the constant $t_{\rm cool}/t_{\rm ff}$ profile
 gives extended multiphase gas for all release radii, extended cold gas condenses out only at radii larger than 80 kpc for 
 the constant entropy profile. The blob starts to fall in much faster after it cools to the stable temperature; this leads to a 
 characteristic `knee' in cases with runaway cooling (e.g., see Figs. \ref{vstime} \& \ref{vstime_hiK}).} \label{fig:lower_mass}
\end{figure}

We have also applied our model (with the same $\beta$ parameter = 0.0244) to a much lower mass Milky-Way-like halo 
with the halo mass of $10^{12} \msun$. Unlike clusters, for such halos cooling is  important even at the virial radius and 
the hot gas density is expected to be much lower (e.g., see Fig. 1 of \citealt{sharma12b}). We have followed the evolution 
of slightly over-dense ($\delta=10^{-5}$) 50 pc blobs starting at rest from various initial radii for two plausible hot gas profiles -- a constant entropy 
($K=80$ keV cm$^2$) core and a constant $t_{\rm cool}/t_{\rm ff}=7$ `core'.  The top panel of Figure \ref{fig:lower_mass} 
shows the entropy and $t_{\rm cool}/t_{\rm ff}$ profiles that we have used. While the constant $\tr$ profile is expected to 
form extended multiphase gas irrespective of the initial radius, the constant entropy profile should give runaway cooling
only at large radii ($\gtrsim$ 80 kpc) where $\tr \lesssim 10$. For the constant entropy profile $\tr$ at 80 kpc is 12.5 and at 60 
kpc it is 20; therefore, runaway cooling  occurs only if $\tr$ slightly exceeds 10. This shows the robustness of our model 
(which uses the same value of the $\beta$ parameter that was fine-tuned for the massive cluster) which gives the threshold 
of $\tr \lesssim 10$ for runaway cooling, even if the halo mass and the ICM profiles are changed drastically.


\item {\em Interplay of local thermal instability and gravity:} Cold gas condenses out much closer in than the location of the minimum of $\tr$. This 
is seen in the idealized simulations (e.g., Fig. 4 in \citealt{sharma12} shows gold gas only extended out to 5 kpc but $\tr<10$ out to 20 kpc for 
the corresponding ICM profile) and in our analytic models (e.g., Fig. \ref{vstime_hiK} shows that the blob released at 20 kpc cools to the stable 
phase only within 5 kpc). This result has implications for the location of high velocity clouds in Milky Way-like galaxies, the farthest of which are 
believed to have condensed out of the hot halo gas. While $\tr \lesssim 10$ even out to the virial radius for $10^{12} \msun$ halos (e.g., 
\citealt{sharma12b}), the cold gas (emitting 21 cm line) may exist well within the halo (see also Fig. \ref{fig:lower_mass}). 
This is consistent with the high velocity clouds detected 
out to 50 kpc from the center of the Andromeda galaxy (\citealt{thilker04}). Note that the cooling time close to the virial radius in Figure 
\ref{fig:lower_mass} can be comparable to the Hubble time and the approximation of a constant halo mass, etc. break down. However, the qualitative
results of our model are expected to hold. Cold gas in galaxy clusters and groups can be pushed farther out 
compared to the predictions of our analytic models and idealized simulations with thermal balance because of the large velocities generated by 
AGN-driven bubbles. 

\citet{malagoli87}, \citet{balbus89} and \citet{binney09}, among others, have investigated the interplay of local thermal instability and gravity. These works have
highlighted the importance of stable stratification in thermally unstable atmospheres, because of which the instability is an over-stability with fast oscillations. 
Using a Lagrangian analysis, \cite{balbus89} have shown that the thermal instability is suppressed in presence of a background cooling flow because the background
conditions change over the same timescale as the instability itself. This was verified in numerical simulations of cooling flows with finite density 
perturbations (see the Appendix section of \citealt{sharma12}) which show that large amplitude density perturbations are required for multiphase 
gas to condense out of the cooling flows. However, in the state of thermal balance (which is strongly favored by observations over the past decade) 
there is no cooling flow and the background state is quasi-static. In such a state even tiny density perturbations can become nonlinear if the cooling time 
of the background ICM is short enough (i.e., $t_{\rm cool}/t_{\rm ff} \lesssim 10$).

The linear response of a strongly-stratified atmosphere in thermal and hydrostatic balance is quite simple (Eq. \ref{eq:LTI}) 
and shows over-stability (see the line marked by circles in Fig. \ref{vstime}) damped nonlinearly by the drag term. 
In  a weakly-stratified spherical ICM, however, a slightly over-dense 
blob becomes further over-dense because of the spherical compression term (see Eq. \ref{eq:linear_sph}; 
spherical compression is unimportant if the entropy scale height is much smaller than the radius) and the heavier blob falls 
inward with a large velocity, at which point the drag force is large and stops the blob from falling in further. Moreover, the spherical compression 
term in Eq. \ref{eq:delDotSoker_sph} becomes weaker as the blob encounters a mass comparable to its own mass. Thus, spherical 
compression is crucial in making it easier for the over-dense blobs to condense out of the ICM.


By constructing strongly stratified artificial ICM profiles, we have verified that the spherical compression term is not effective when 
the entropy scale height is much smaller than the blob release radius (i.e., if the atmosphere is strongly stratified; see Eq. \ref{eq:linear_sph}). 
We can also understand the dependence of the critical $\tr$ on the background entropy profile in cylindrical geometry from Eq. \ref{eq:linear_sph}
(as mentioned in section \ref{sec:sph_comp}, the critical value varies from 2.5 to 5 for typical cool cluster entropy profiles). 
For weak entropy stratification of the ICM the spherical compression term in Eq. \ref{eq:linear_sph} overwhelms
the oscillatory $N^2$ term. For cylindrical geometry, however, the compression term is half of its spherical value ($-g/r$ instead of $-2g/r$ in 
Eq. \ref{eq:linear_sph}) and even weak stratification of the ICM decreases the critical $\tr$ for condensation. Therefore, in cylindrical geometry 
the critical $\tr$ is larger if the ICM is weakly stratified (e.g., $K \propto r^0$) and smaller if it is even slightly stratified (e.g., $K \propto r^{1.1}$).

Stratification is not expected to be very strong for astrophysical coronae, which by definition, are close to the virial temperature. Thus, 
our $\tr \lesssim 10$ criterion for the condensation of extended cold gas is expected to be valid in most astrophysical coronae, ranging from 
the solar corona to hot accretion flows (e.g., see \citealt{sharma13,das13,gaspari13}). 

\item {\em Extended multiphase gas and cold feedback:} The accretion of {\em cold} gas by the central massive black holes in cluster cores is essential for 
AGN feedback to close the {\em globally stable} feedback loop. Hot accretion rate via Bondi accretion is small and not as sensitively dependent on the 
ICM density ($\propto n$) as radiative cooling ($\propto n^2$). Therefore, accretion in the hot phase alone seems incapable of globally balancing 
radiative cooling. Moreover, the Bondi accretion rate is about two orders of magnitude smaller than the multiphase mass cooling rate close to 
10 kpc, which should be of order the black hole accretion rate (e.g., see \citealt{gaspari13}). Moreover, as discussed in detail in section 5.1 
of \citet{sharma12}, the cold feedback model may also naturally account for the observed correlation between the estimated Bondi accretion rate and 
the jet power.

Our phenomenological model tries to provide a physical basis for the condensation of cold gas in the ICM. In particular, it tries to explain the large quantitative
difference in the condition for the condensation of multiphase gas in a plane-parallel and a spherical atmosphere; namely, $\tr \lesssim 1$ versus 10 for the 
condensation of cold gas starting  from small perturbations. 
Thus, we have provided a firmer footing to the cold
feedback paradigm by extending PS05's model to agree with observations and numerical simulations. We also argue that the characteristic shape of $\tr$ profile 
with a minimum at $\sim 10$ kpc (rather than right at the center) is very crucial. Because of this, a large amount of gas far away from the sphere of influence 
of the supermassive black hole (unlike Bondi accretion) can condense out  and episodically boost the accretion rate (due to cold gas) by a large amount. 
This can rather effectively stop the cooling flow and make  $\tr > 10$ throughout, and the condensation of cold gas is suppressed. The core can form again 
because of a feeble accretion rate in the hot mode, and once $\tr \lesssim 10$ again, the core can suffer heating due to AGN jets/cavities and the global 
cycle continues. The presence of extended cold gas also agrees with the observations of \citet{mcdonald10} and others, which show extended cold 
gas at 10s of kpc where $\tr \lesssim 10$ (e.g., see Fig. 11 of \citealt{mccourt12}).

\end{enumerate}

\section*{Acknowledgements}
The authors thank Noam Soker for his comments on the paper. AS thanks the Indian Academy of Sciences, Indian National Science Academy 
and the National Academy of Sciences India for awarding the Summer Research Fellowship 2012 with grant number PHYS1231. This work is 
partly supported by the DST-India grant no. Sr/S2/HEP-048/2012.

\bibliographystyle{mn2e}
\bibliography{bibtex}

\begin{thebibliography}{38}
\expandafter\ifx\csname natexlab\endcsname\relax\def\natexlab#1{#1}\fi

\bibitem[{{Balbus} \& {Soker}(1989)}]{balbus89}
{Balbus} S.~A., {Soker} N., 1989, \apj, 341, 611

\bibitem[{{Banerjee} \& {Sharma}(2014)}]{banerjee14}
{Banerjee} N., {Sharma} P., 2014, \mnras, 443, 687

\bibitem[{{Binney}, {Nipoti} \& {Fraternali}(2009){Binney}, {Nipoti}, \&
  {Fraternali}}]{binney09}
{Binney} J., {Nipoti} C., {Fraternali} F., 2009, \mnras, 397, 1804

\bibitem[{{Burkert} \& {Lin}(2000)}]{burkert00}
{Burkert} A., {Lin} D.~N.~C., 2000, \apj, 537, 270

\bibitem[{{Cavagnolo} {et~al}\mbox{.}(2009){Cavagnolo}, {Donahue}, {Voit}, \&
  {Sun}}]{accept09}
{Cavagnolo} K.~W., {Donahue} M., {Voit} G.~M., {Sun} M., 2009, \apjs, 182, 12

\bibitem[{{Churazov} {et~al}\mbox{.}(2001){Churazov}, {Br{\"u}ggen}, {Kaiser},
  {B{\"o}hringer}, \& {Forman}}]{chu01}
{Churazov} E., {Br{\"u}ggen} M., {Kaiser} C.~R., {B{\"o}hringer} H., {Forman}
  W., 2001, \apj, 554, 261

\bibitem[{{Churazov} {et~al}\mbox{.}(2008){Churazov}, {Forman}, {Vikhlinin},
  {Tremaine}, {Gerhard}, \& {Jones}}]{churazov08}
{Churazov} E., {Forman} W., {Vikhlinin} A., {Tremaine} S., {Gerhard} O.,
  {Jones} C., 2008, \mnras, 388, 1062

\bibitem[{{Das} \& {Sharma}(2013)}]{das13}
{Das} U., {Sharma} P., 2013, \mnras, 435, 2431

\bibitem[{{Dubois} {et~al}\mbox{.}(2011){Dubois}, {Devriendt}, {Teyssier}, \&
  {Slyz}}]{dubois11}
{Dubois} Y., {Devriendt} J., {Teyssier} R., {Slyz} A., 2011, \mnras, 417, 1853

\bibitem[{Edge(2001)}]{edge2001}
Edge A., 2001, Monthly Notices of the Royal Astronomical Society, 328, 762

\bibitem[{{Fabian}(1994)}]{fabian94}
{Fabian} A.~C., 1994, \araa, 32, 277

\bibitem[{{Field}(1965)}]{field65}
{Field} G.~B., 1965, \apj, 142, 531

\bibitem[{{Gaspari}, {Ruszkowski} \& {Oh}(2013){Gaspari}, {Ruszkowski}, \&
  {Oh}}]{gaspari13}
{Gaspari} M., {Ruszkowski} M., {Oh} S.~P., 2013, \mnras, 432, 3401

\bibitem[{{Gaspari}, {Ruszkowski} \& {Sharma}(2012){Gaspari}, {Ruszkowski}, \&
  {Sharma}}]{gaspari12}
{Gaspari} M., {Ruszkowski} M., {Sharma} P., 2012, \apj, 746, 94

\bibitem[{{Leccardi} \& {Molendi}(2008)}]{leccardi08}
{Leccardi} A., {Molendi} S., 2008, \aap, 487, 461

\bibitem[{{Li} \& {Bryan}(2013)}]{li14}
{Li} Y., {Bryan} G.~L., 2013, ArXiv e-prints

\bibitem[{{Malagoli}, {Rosner} \& {Bodo}(1987){Malagoli}, {Rosner}, \&
  {Bodo}}]{malagoli87}
{Malagoli} A., {Rosner} R., {Bodo} G., 1987, \apj, 319, 632

\bibitem[{{McCourt} {et~al}\mbox{.}(2012){McCourt}, {Sharma}, {Parrish}, \&
  {Quataert}}]{mccourt12}
{McCourt} M., {Sharma} P., {Parrish} I.~J., {Quataert} E., 2012, \mnras, 419,
  3319

\bibitem[{{McDonald} {et~al}\mbox{.}(2010){McDonald}, {Veilleux}, {Rupke}, \&
  {Mushotzky}}]{mcdonald10}
{McDonald} M., {Veilleux} S., {Rupke} D.~S.~N., {Mushotzky} R., 2010, \apj,
  721, 1262

\bibitem[{{McNamara} \& {Nulsen}(2007)}]{mcnamara2007}
{McNamara} B.~R., {Nulsen} P.~E.~J., 2007, \araa, 45, 117

\bibitem[{{Mohr}(1999)}]{mo99}
{Mohr} J.~J., 1999, \apj, 517, 627

\bibitem[{{Navarro}, {Frenk} \& {White}(1996){Navarro}, {Frenk}, \&
  {White}}]{nfw96}
{Navarro} J.~F., {Frenk} C.~S., {White} S.~D.~M., 1996, \apj, 463, 563

\bibitem[{O'Dea {et~al}\mbox{.}(2008)O'Dea, Baum, Privon, Noel-Storr, Quillen,
  Zufelt, Park, Edge, Russell, Fabian, Donahue, Sarazin, McNamara, Bregman, \&
  Egami}]{odea2008}
O'Dea C.~P. {et~al.}, 2008, The Astrophysical Journal, 681, 1035

\bibitem[{Peterson {et~al}\mbox{.}(2003)Peterson, Kahn, Paerels, Kaastra,
  Tamura, Bleeker, Ferrigno, \& Jernigan}]{peterson2003}
Peterson J.~R., Kahn S.~M., Paerels F. B.~S., Kaastra J.~S., Tamura T., Bleeker
  J. A.~M., Ferrigno C., Jernigan J.~G., 2003, The Astrophysical Journal, 590,
  207

\bibitem[{{Pizzolato} \& {Soker}(2005)}]{ps05}
{Pizzolato} F., {Soker} N., 2005, \apj, 632, 821

\bibitem[{{Pizzolato} \& {Soker}(2010)}]{ps10}
{Pizzolato} F., {Soker} N., 2010, \mnras, 408, 961

\bibitem[{{Quataert}(2008)}]{quataert08}
{Quataert} E., 2008, \apj, 673, 758

\bibitem[{{{Rybicki}, G.~B. and {Lightman}, A.~P.}(2004)}]{ryb+light}
{{Rybicki}, G.~B. and {Lightman}, A.~P.}, 2004, {Radiative Processes in
  Astrophysics}. {Wiley-VCH}, p. 155

\bibitem[{{Sharma}(2013)}]{sharma13}
{Sharma} P., 2013, in Astronomical Society of India Conference Series, Vol.~9,
  Astronomical Society of India Conference Series, pp. 27--31

\bibitem[{{Sharma} {et~al}\mbox{.}(2009){Sharma}, {Chandran}, {Quataert}, \&
  {Parrish}}]{sharma09}
{Sharma} P., {Chandran} B.~D.~G., {Quataert} E., {Parrish} I.~J., 2009, in
  American Institute of Physics Conference Series, Vol. 1201, American
  Institute of Physics Conference Series, {Heinz} S., {Wilcots} E., eds., pp.
  363--370

\bibitem[{{Sharma} {et~al}\mbox{.}(2012{\natexlab{a}}){Sharma}, {McCourt},
  {Parrish}, \& {Quataert}}]{sharma12b}
{Sharma} P., {McCourt} M., {Parrish} I.~J., {Quataert} E., 2012{\natexlab{a}},
  \mnras, 427, 1219

\bibitem[{{Sharma} {et~al}\mbox{.}(2012{\natexlab{b}}){Sharma}, {McCourt},
  {Quataert}, \& {Parrish}}]{sharma12}
{Sharma} P., {McCourt} M., {Quataert} E., {Parrish} I.~J., 2012{\natexlab{b}},
  \mnras, 420, 3174

\bibitem[{{Sharma}, {Parrish} \& {Quataert}(2010){Sharma}, {Parrish}, \&
  {Quataert}}]{sharma10}
{Sharma} P., {Parrish} I.~J., {Quataert} E., 2010, \apj, 720, 652

\bibitem[{{Sutherland} \& {Dopita}(1993)}]{sutherland+dopita+93}
{Sutherland} R.~S., {Dopita} M.~A., 1993, \apjs, 88, 253

\bibitem[{{Thilker} {et~al}\mbox{.}(2004){Thilker}, {Braun}, {Walterbos},
  {Corbelli}, {Lockman}, {Murphy}, \& {Maddalena}}]{thilker04}
{Thilker} D.~A., {Braun} R., {Walterbos} R.~A.~M., {Corbelli} E., {Lockman}
  F.~J., {Murphy} E., {Maddalena} R., 2004, \apjl, 601, L39

\bibitem[{{Tozzi} \& {Norman}(2001)}]{tozzi+norman+2001}
{Tozzi} P., {Norman} C., 2001, \apj, 546, 63

\bibitem[{{Wagh}, {Sharma} \& {McCourt}(2014){Wagh}, {Sharma}, \&
  {McCourt}}]{wagh14}
{Wagh} B., {Sharma} P., {McCourt} M., 2014, \mnras, 439, 2822

\bibitem[{{Werner} {et~al}\mbox{.}(2009){Werner}, {Zhuravleva}, {Churazov},
  {Simionescu}, {Allen}, {Forman}, {Jones}, \& {Kaastra}}]{werner09}
{Werner} N., {Zhuravleva} I., {Churazov} E., {Simionescu} A., {Allen} S.~W.,
  {Forman} W., {Jones} C., {Kaastra} J.~S., 2009, \mnras, 398, 23

\end{thebibliography}

\label{lastpage}

\end{document}